# Automated detection of gibbon calls from passive acoustic monitoring data using convolutional neural networks in the 'torch for R' ecosystem

Dena J. Clink[1*], Jinsung Kim[1], Hope Cross-Jaya[1], Abdul Hamid Ahmad[2], Moeurk Hong[3], Roeun Sala[3], Hélène Birot[3], Cain Agger[4], Thinh Tien Vu[5], Hoa Nguyen Thi[6], Thanh Nguyen Chi[7], and Holger Klinck[1]

[1] K. Lisa Yang Center for Conservation Bioacoustics, Cornell Lab of Ornithology, Cornell University, Ithaca, NY, USA.
[2] Institute for Tropical Biology and Conservation, Universiti Malaysia Sabah (UMS), Kota Kinabalu, Sabah, Malaysia
[3] Jahoo, Angdoung Kraleung Village, Sen Monorom Orang, Mondulkiri, Mondulkiri Province, Cambodia
[4] Wildlife Conservation Society, Cambodia, 21, Street 21, Sangkat Tonle Bassac, Phnom Penh, Cambodia.
[5] Department of Wildlife, Vietnam National University of Forestry, Xuan Mai, Chuong My, Ha Noi, Vietnam.
[6] The Institute for Tropical Biodiversity and Forestry, Xuan Mai, Chuong My, Ha Noi, Vietnam.
[7] Bac Giang Agro-Forestry University, Bich Son Commune, Viet Yen District, Bac Giang Province, Vietnam

**Running title:** Gibbon calls using 'torch for R'

**\* Corresponding author email:**
Email: dena.clink@cornell.edu

**Abstract:**
Automated detection of acoustic signals is crucial for effective monitoring of sound-producing animals and their habitats across ecologically relevant spatial and temporal scales. Recent advances in deep learning have made these approaches more accessible. However, few deep learning approaches can be implemented natively in the R programming environment; approaches that run natively in R may be more accessible for ecologists. The 'torch for R' ecosystem has made deep learning with convolutional neural networks accessible for R users. Here, we evaluate a workflow for the automated detection and classification of acoustic signals from passive acoustic monitoring (PAM) data. Our specific goals include: 1) present a method for automated detection of gibbon calls from PAM data using the 'torch for R' ecosystem; 2) conduct a series of benchmarking experiments and compare the results of six CNN architectures; and 3) investigate how well the different architectures perform on datasets of the female calls from two different gibbon species: the northern grey gibbon (*Hylobates funereus*) and the southern yellow-cheeked crested gibbon (*Nomascus gabriellae*). We found that the highest-performing architecture depended on the species and test dataset. We successfully deployed the top-performing model for each gibbon species to investigate spatial variation in gibbon calling behavior across two grids of autonomous recording units in Danum Valley Conservation Area, Malaysia and Keo Seima Wildlife Sanctuary, Cambodia. The fields of deep learning and automated detection are rapidly evolving, and we provide the methods and datasets as benchmarks for future work.

**Introduction:**

*Passive acoustic monitoring*

Terrestrial applications of passive acoustic monitoring (PAM) – which utilizes autonomous acoustic recording units (ARUs) – have increased dramatically in recent years [reviewed in (Sugai et al., 2019)]. The use of autonomous recording units allows for monitoring at spatial and temporal scales that are generally not achievable using human observers (Gibb et al., 2018). PAM has been used to investigate vocal behavior (Clink et al., 2020), model occurrence probability in the landscape (Vu & Tran, 2019), and for automated detection/classification of calls from long-term PAM recordings (Clink et al., 2023; Dufourq et al., 2021). However, the use of PAM often results in terabytes of acoustic data that require post-processing to obtain useful information about the signals of interest. A major bottleneck in using PAM for monitoring populations is related to extracting relevant information from long-term acoustic recordings (Tuia et al., 2022), and listening to the recordings or manual annotation is time- and cost-prohibitive. Therefore, identifying effective automated approaches is critical for the effective use of PAM.

*Automated detection/classification*

Numerous approaches have been developed for the automated detection and classification of animal sounds from terrestrial PAM data. In the context of PAM, classification can be defined as the assignment of each observation (in this case sound clip) to a respective class (e.g., species, individual) and detection is the use of a sliding window approach to identify signals of interest from background noise in long recordings (Stowell, 2022). Some of the earlier approaches for automated detection include spectrogram cross-correlation (Katz et al., 2016), or combining band-limited energy summation with a subsequent classifier (Clink et al., 2023; Kalan et al., 2015; Ross, 2013). Recent advances in deep learning have revolutionized image and speech recognition (LeCun et al., 2015), with important cross-over for bioacoustics and the analysis of PAM data. One of the most important innovations was applying convolutional neural network (CNN) architecture, that has been hugely successful for image classification, to audio data (Hershey et al., 2017). There has been a huge increase in the use of deep learning for the automated detection of signals in PAM data in recent years [reviewed in (Stowell, 2022)]. This includes many deep learning applications for terrestrial PAM data, including anurans (LeBien et al., 2020), birds (Kahl et al., 2021; Stowell et al., 2019), bats (Aodha et al., 2018), and primates (Dufourq et al., 2020; Ravaglia et al., 2023).

*Convolutional neural networks*

In the most fundamental form, deep learning for classification problems maps the input (e.g., spectrogram image) to the label (e.g., gibbon) via a series of layered transformations so that inputs can be correctly matched to their associated targets (Wani et al., 2020). Traditional approaches to machine learning for acoustic data relied heavily on feature engineering, as early machine learning algorithms required a reduced set of representative features, such as features estimated from the spectrogram including low frequency, high frequency, and duration of the signal. Deep learning does not require feature engineering (Stevens et al., 2020). Convolutional neural networks (CNNs) are useful for processing data that have a 'grid-like topology', such as image or spectrogram data that can be considered a 2-dimensional grid of pixels (Goodfellow et al., 2016). The 'convolutional' layer learns the feature representations of the inputs; these convolutional layers consist of a set of filters which are fundamentally two-dimensional matrices of numbers with the primary parameter being the number of filters (Gu et al., 2018). However, if training data are scarce, training CNNs from scratch may lead to overfitting as representations of images tend to be large with many variables (LeCun et al., 1995).

*Deep learning and PAM*

Transfer learning is an approach wherein the architecture of a pretrained CNN (which is generally trained on a large dataset) is applied to a new classification problem (Dufourq et al., 2022). For example, CNNs trained on the ImageNet dataset of > 1 million images (Deng et al., 2009) such as ResNet have been applied to automated detection/classification of primate and bird species from PAM data (Dufourq et al., 2022; Ruan et al., 2022). Transfer learning in computer vision applications either: 1) retains the feature extraction or embedding layers by freezing the weights, and modifies the last few classification layers to be trained for a new classification task, or 2) initializes the model with pre-trained weights but allows some or all of the layers to be fine-tuned during training (Dufourq et al., 2022). Transfer learning has been shown to outperform CNNs trained with random initial weights in cases where training data are scarce (Tan et al., 2018). Transfer learning is particularly appropriate when there is a paucity of training data (Weiss et al., 2016), such as common in PAM data.

Transfer learning has been applied to PAM data in a variety of ways, with one of the most notable differences in approaches being related to the type of data used to train the model. It is common practice to compare the performance of a variety of different architectures on the same dataset, as oftentimes there are no *a priori* reasons to expect that one architecture will perform better than another, and performance may vary across signal types and architectures. Dufourq et al. (2022) compared different transfer learning architectures that were pretrained on the ImageNet dataset for the automated detection of Hainan gibbons (*Nomascus hainanus*), black-and-white ruffed lemurs (*Varecia variegate)*, Thyolo alethe (*Chamaetylas choloensis)*, and the Pin-tailed whydah (*Vidua macroura*). They found that performance was dependent on model configuration, but pre-trained ResNet152V2 had consistently high performance with relatively few training samples (25 samples) and achieved an F1 score of ~0.8 for gibbons and lemurs.

The BEnchmark of ANimal Sounds (BEANS) compared different deep learning and non-deep learning algorithms on 12 datasets (Hagiwara et al., 2022), including a comparison of pretrained ResNets (trained on the ImageNet) dataset to the VGGish (Hershey et al., 2017; Simonyan & Zisserman, 2014) model that was pretrained on audio from the YouTube dataset (Gemmeke et al., 2017). They found that performance was variable across datasets and tasks (e.g. classification versus detection), and that the VGGish model generally performed the best, followed by the pretrained ResNets. For the Hainan gibbon dataset they found that the pretrained ResNET 18 architecture performed best, but with a relatively low performance of mean average precision = 0.3. For a few of the classification tasks a non-deep learning algorithm – support vector machine (Cortes & Vapnik, 1995) – performed best. Another approach used models pretrained on global datasets of bird vocalizations to extract embeddings and a subsequently trained new classifier [BirdNET (Kahl et al., 2021) and Perch (Ghani et al., 2023)]. The authors compared these models to those pretrained on the YouTube and AudioSet data sets (Ghani et al., 2023) and found that BirdNET and Perch substantially outperformed the other models.

*The 'torch for R' ecosystem*

The two most popular open-source programming languages for ecological applications are R and Python (Scavetta & Angelov, 2021). Python has surpassed R in terms of overall popularity, but R remains an important language for the life sciences (Lawlor et al., 2022). 'Keras' (Chollet & others, 2015), 'PyTorch' (Paszke et al., 2019) and 'Tensorflow' (Martín Abadi et al., 2015) are some of the more popular neural network libraries; these libraries were all initially developed for the Python programming language. Until recently, deep learning implementations in R relied on the 'reticulate' package which served as an interface to Python (Ushey et al., 2022). However, the recent release of the 'torch for R' ecosystem provides a framework based on 'PyTorch' that runs natively in R and has no dependency on Python (Falbel & Luraschi, 2023). Running natively in R means more straightforward installation, and higher accessibility for users of the R programming environment. Torch for R provides GPU acceleration, like the Python implementation, however, it does not have all the functionality of the Python version as it is actively under development (as of 2025).

*Gibbons & PAM*

Gibbons (family Hylobatidae) are small apes that are found throughout Southeast Asia, and most of the ~20 species of gibbons are endangered or critically endangered (IUCN, 2022). All gibbon species have species- and sex-specific loud calls that can be heard at distances of > 1 km (Mitani, 1985), which makes them good candidates for PAM. Traditional methods of monitoring gibbon populations have relied on human observers (Brockelman & Srikosamatara, 1993; Kidney et al., 2016), but these approaches are time- and labor-intensive. A few gibbon species have been studied using PAM, including the Hainan gibbon, *Nomascus hainanus* (Dufourq et al., 2020), cao vit gibbon, *Nomascus nasutus* (Wearn et al., 2024), Northern grey gibbon, *Hylobates funereus* (Clink et al., 2020, 2023), western black crested gibbon, *N. concolor* (Zhong et al., 2021) and the southern yellow-cheeked crested gibbon, *N. gabriellae* (Vu & Tran, 2019). To-date, the automated detection/classification of gibbon signals has been done for four gibbon species, western black-crested gibbon (*Nomascus concolor*) (Zhou et al., 2023), Hainan gibbons (Dufourq et al., 2020), Bornean white-bearded gibbon (*H. albibarbis*) (Owens et al., 2024), and Northern grey gibbons (Clink et al., 2023). However, the increasing accessibility of autonomous recording units and analytical approaches means that more gibbon species will be soon added to the list

      Effective automated detection approaches could help improve monitoring and conservation efforts of gibbons using PAM. For example, previous work applied occupancy modeling to PAM data, but the acoustic data were analyzed manually (Vu et al., 2023; Vu & Tran, 2020, 2019). Combining PAM, automated detection, and occupancy modeling can provide information on spatial and temporal scales that are difficult to achieve if analyzing the data manually (Wood et al., 2024). PAM in combination with automated detection can also be used to monitor individual- and group-level dynamics of gibbons (Wang et al., 2024); these authors used a traditional machine learning approach. Automated detection could also improve our ability to understand the behavioral ecology of gibbons. For example, a PAM study investigating

the impacts of rain on gibbon vocal behavior used manual annotation instead of automated detection, but the use of an automated detector would have made the analysis more efficient and scalable (Clink et al., 2020).

*Objectives*

The 'torch for R' ecosystem is under rapid development, and there are existing methods and tutorials for image classification using pretrained CNNs (Keydana, 2023). The goals of the current study include: 1) present a method for automated detection of gibbon calls from PAM data using the 'torch for R' ecosystem; 2) compare the results for six CNN architectures AlexNet (Krizhevsky et al., 2017), VGG16, VGG19 (Simonyan & Zisserman, 2014), ResNet18, ResNet50, and ResNet152 (He et al., 2016) pretrained on the 'ImageNet' dataset (Deng et al., 2009); and 3) investigate how well the different architectures perform on datasets of the female calls from two different gibbon species: the northern grey gibbon (*Hylobates funereus*) and the southern yellow-cheeked crested gibbon (*Nomascus gabriellae*). These species are not sympatric; however, we predicted that including more training samples from different classes would improve model performance by increasing the diversity of sounds seen during training, therefore allowing the models to generalize more effectively.

We also conducted a series of benchmarking experiments to: 1) investigate how stochasticity in the training process influences performance metrics; 2) compare the performance of fine-tuned models with that of models where feature extraction layers were kept frozen and only the classification layers were retrained; 3) explore the influence of data augmentation on model performance metrics; and 4) compare this approach with current state-of-the art bioacoustics models. Our team recently benchmarked the approach presented here with other models, including BirdNET (Kahl et al., 2021) and Koogu (Madhusudhana, 2023) for southern yellow-cheeked crested gibbons, and we found comparable performance when number of training samples was > 200 (Clink et al., 2024). To expand on our previous findings, we also include a comparison of the 'torch for R' approach with that of BirdNET for both gibbon species in the present study.

We evaluated the performance of both binary classification and multi-class classification approaches, as we wanted to see if including more gibbon training samples (regardless of the species) would improve performance. We then deployed the top-performing model for each gibbon species to investigate spatial patterns of variation in gibbon calling behavior across two grids of autonomous recording units in Danum Valley Conservation Area, Malaysia and Keo Seima Wildlife Sanctuary, Cambodia. For both gibbon species, we test the generalizability of the models on data collected from a different site than the training data; this is considered best practice for machine learning (Stowell, 2022).

**Materials and Methods:**

*Acoustic data collection*

Acoustic data used for training were collected using Swift or SwiftOne (K. Lisa Yang Center for Conservation Bioacoustics, Cornell Lab of Ornithology, Cornell University, USA) autonomous recording units at two locations in Southeast Asia. The first recording location for training data was Danum Valley Conservation Area (Danum Valley) in Sabah, Malaysia; the gibbon species present here is the northern grey gibbon (hereafter grey gibbons). Acoustic data at Danum Valley were collected using Swift autonomous recording units from February to April 2018 at 40 dB gain, 16 kHz sample rate, and 16-bit resolution on a 24-hr continuous recording schedule. Recordings were saved as 2-hour Waveform Audio File Format (.wav) files, and each file was ~230 MB in size. Danum Valley is considered 'aseasonal' and lacks monsoons that are typical of other areas in Southeast Asia (Walsh & Newbery, 1999). The second recording location, which we used for test data to evaluate the generalizability of model performance, was Maliau Basin Conservation Area which is ~ 100 km from Danum Valley. Acoustic data from Maliau Basin were collected in August 2019 using Swifts at a 40 dB gain, 48 kHz sampling rate, and 16-bit resolution on a 24-hr continuous recording schedule. Recordings were saved as 40-min files that were approximately ~230 MB in size. The sensitivity of the Swift microphones was −44 (+/−3) dB re 1 V/Pa with ADC clipping level of -/+ 0.9V.

The second recording location for training data was Jahoo, Mondulkiri Province, Cambodia, and the southern yellow-cheeked crested gibbons are found here (hereafter crested gibbons). Acoustic data collection was done using SwiftOnes with a 32 dB gain, 32 kHz sample rate, 16-bit resolution, with 1-hour .wav files at ~230 MB size on a 24-hr continuous recording schedule. The sensitivity of the SwiftOne microphones was −24 (+/−3) dB re 1 V/Pa with ADC clipping level of -/+ 0.9V. Training data was taken from recordings collected from March to May 2022. Mondulkiri province experiences a distinct wet season driven by the monsoon from May to September each year. The test dataset for crested gibbons comes from Dakrong Nature Reserve in Vietnam and contains recordings of a recently recognized distinct gibbon species (Van Chuong et al., 2018), the Northern buff-cheeked gibbons (*N. annamensis*), which was originally classified as *N. gabriellae*. Data from this site were collected using a modified smartphone at 0 gain, 16 kHz sample rate, 16-bit resolution, on a 24-hr continuous recording schedule, and saved as 1-hour .wav files (Vu et al., 2023). See Figure 1 for a map of the recording locations in Malaysia, Cambodia, and Vietnam, and see Table 1 for a summary of sample size for different datasets. GPS locations of all recording locations are available on Github (see data availability statement).

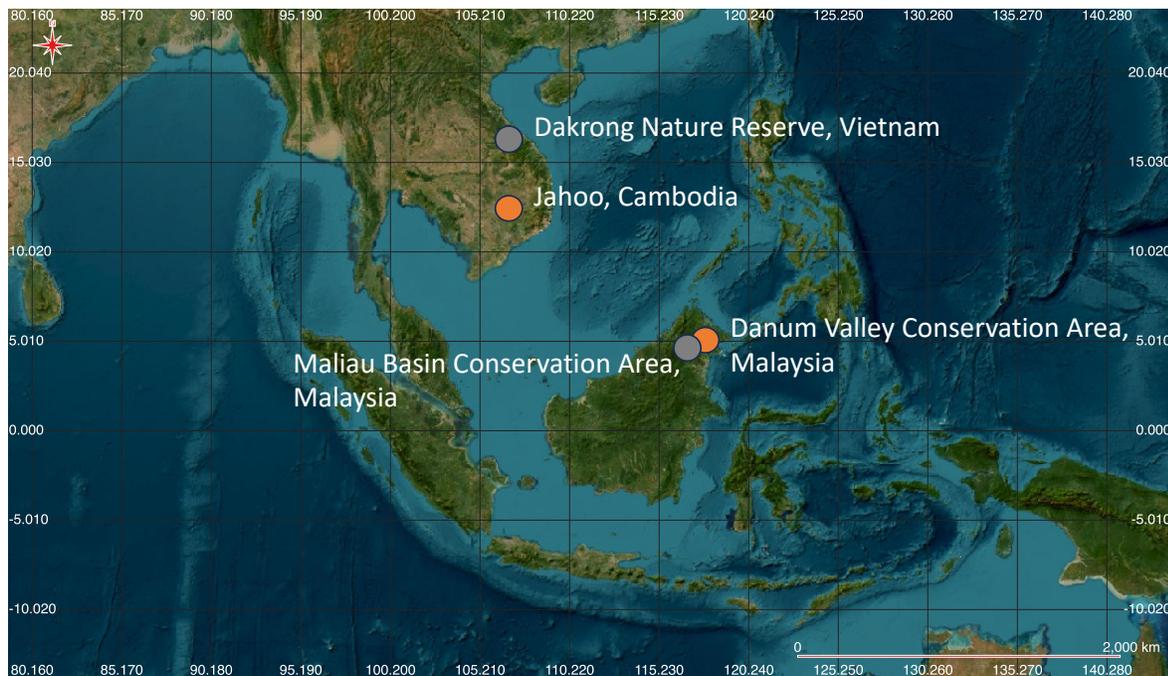

**Figure 1. Map of recording locations in the present study.** Northern grey gibbons (*Hylobates funereus*) are found in Danum Valley and Maliau Basin, southern yellow cheeked-crested gibbons (*Nomascus gabriellae*) are found in Jahoo, Cambodia, and northern buff-cheeked gibbons (*Nomascus annamensis*) are found in Dakrong Nature Reserve, Vietnam. Recording locations used for training datasets are indicated in orange and test datasets in grey. Map was created using QGIS (QGIS.org, %Y. QGIS Geographic Information System. QGIS Association. http://www.qgis.org).

**Table 1. Summary of sample size for training and test sets used in the present study.** For Danum Valley and Jahoo test datasets, the models were deployed over the entire sound files, so the number of noise samples is substantially higher than the other datasets.

| Dataset | Country | Species | Number of Gibbon samples | Number of Noise samples | Sample rate |
|---|---|---|---|---|---|
| Danum Valley Conservation Training Data | Malaysia | Grey gibbon | 502 | 2,254 | 16 kHz |
| Jahoo Training Data | Cambodia | Crested gibbon | 213 | 2,130 | 32 kHz |
| Maliau Basin Test Data | Malaysia | Grey gibbon | 147 | 81 | 48 kHz |
| Dakrong Nature Reserve Test Data | Vietnam | Buff cheeked gibbon | 45 | 173 | 16 kHz |
| Danum Valley Test Data | Malaysia | Grey gibbon | 383 | 2,905 | 16 kHz |
| Jahoo Test Data | Cambodia | Crested gibbon | 296 | 5,684 | 32 kHz |

*Training data preparation*

Acoustic samples used for the training data were prepared slightly differently for each species, as the data were compiled from two separate projects. For grey gibbons, we randomly selected 500 hours of acoustic recordings from 06:00 to 10:00 LT, as this is when grey gibbons are most likely to call at this site (Clink et al., 2020). We used a band-limited energy detector implemented using the 'DetectBLED' function in the 'gibbonR' package (Clink & Klinck, 2019) to isolate sound events in the 0.5 – 1.6 kHz frequency range. One analyst (DJC) created spectrogram images of the clips using the 'seewave' R package (Sueur et al., 2008), and assigned each sound event as either a female "gibbon" call or a catch-all "noise" category; this resulted in 502 female grey gibbon calls and 2,254 noise events. See Clink et al. (2023) for details on the band-limited energy detector implementation and settings. For crested gibbons, we randomly selected 789 hours of recordings and two analysts (JK and HCJ) manually annotated all instances of female gibbon calls using spectrograms created in Raven Pro 1.6.3 (K. Lisa Yang Center for Conservation Bioacoustics at the Cornell Lab of Ornithology, 2023) setting window size = 2,400 samples, contrast = 70, and brightness = 65; all other settings were the default. We also indicated whether the calls were high-, medium- or low-quality, and omitted the low-quality calls from our training dataset. We classified calls as high-quality if they had visible harmonics, indicating the calling animals were close to the recording unit. Medium-quality calls had few visible harmonics but still had clear visible structure stereotypical of the gibbon female call. Low quality calls either had low signal-to-noise ratio (< 10 dB signal-to-noise ratio) or exhibited substantial overlap with another non-gibbon signal. We divided 1-hr recordings into 12-s windows with 6-s overlap and considered all windows that contained at least 80% of the gibbon call as a positive signal. This resulted in 213 female gibbon calls for training. To create a noise class, we used the band-limited energy detector in the 0.5 – 3.0 kHz frequency range on 20 of the randomly selected files described above that were confirmed to not have gibbon calls present. This resulted in over 10,000 noise clips, so to create a more comparable dataset to the grey gibbons we randomly selected 2,130 noise clips to use for training.

*Test data preparation*

To create test datasets for generalizability, DJC and TTV manually annotated spectrograms indicating the start and stop times of gibbon female calls using Raven Pro 1.6.3 and the default settings. To report the generalizability of our models we annotated nine randomly selected files from Maliau Basin Conservation Area, and for Dakrong Nature Reserve, Vietnam, we included five annotated files that were known to contain gibbon calls. It is important to note that this approach leads to a different class balance than randomly selecting calls, with a larger number of gibbon samples relative to the noise class, which means that performance metrics are likely higher than if we used completely random recordings. To report the final performance of the models, we annotated 10 randomly selected 2-hour files from the wide array in Danum Valley Conservation Area, Malaysia (same recorder settings and time period as above but different ARU locations within Danum Valley), and for the wide array in Jahoo, Cambodia (same recorder settings as above, with recordings from July to November 2022 at from a grid approximately 2-km away from the training grid) we annotated 10 randomly selected 1-hour. To create test datasets for final model performance evaluation that reflected 'real-world' conditions for automated detection, we annotated all gibbon calls that were visible in the spectrogram and could be confirmed aurally to be gibbon calls, even if the full species-specific structure of the calls was not visible. We divided the test recordings into 12-s clips with 6-s hop size or overlap and considered a 12-s clip to be a positive (i.e., containing a gibbon call) if the start time of the clip fell within ±6 seconds of the start of any annotated gibbon call. We then manually verified all clips to ensure that they were assigned to the correct category. See Table 1 for summary of sample size for the different training and test datasets.

*Spectrogram image preparation*

We used the 'spectro' function in the 'seewave' R package (Sueur et al., 2008) to create spectrograms to input into the models. For all recordings that did not have 16 kHz sample rate we down-sampled the clips before converting to spectrograms to ensure comparable time and frequency resolution. We did not filter before down-sampling, however visual inspection of spectrogram images did not show any signs of aliasing for the gibbon frequency range. We used the default window size and color scheme, but removed all axis labels and specified a frequency range of 0.5 – 1.6 kHz for grey gibbons and 0.5 – 3.0 kHz for crested gibbons. We saved all spectrogram images as '.jpg' files using the 'jpeg' function in base R. See Figure 2 for representative spectrogram images of the three gibbon species, along with some representative noise images, included in the present study.

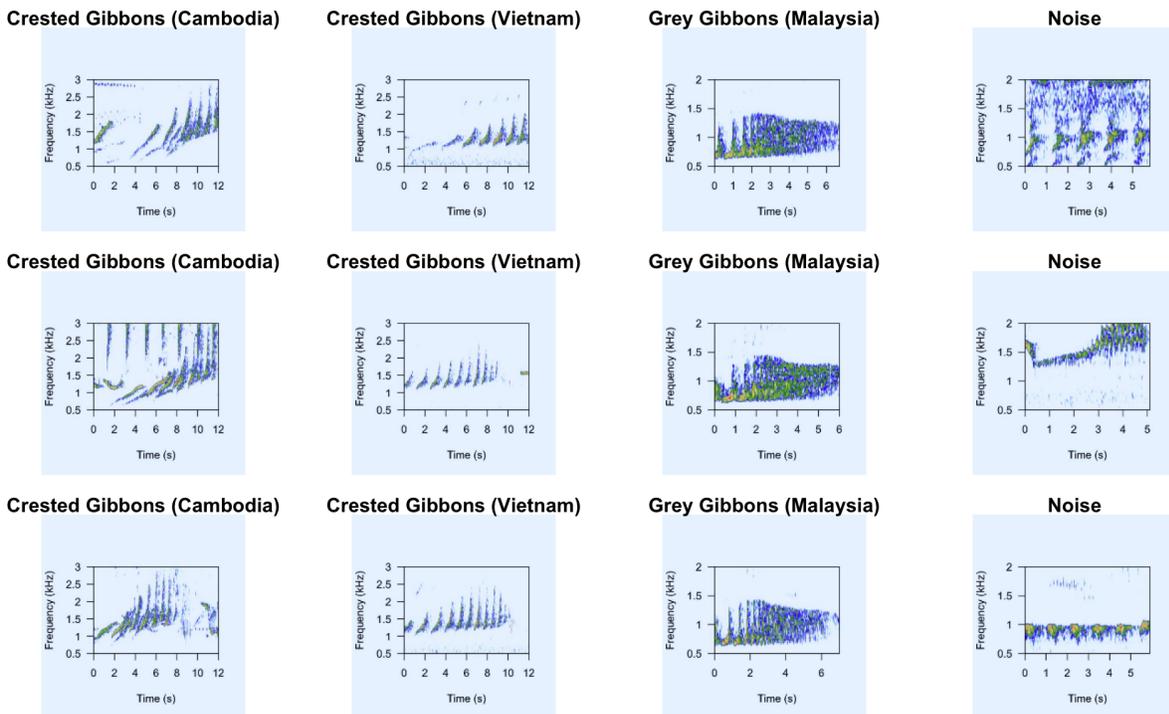

**Figure 2. Representative spectrogram images of female calls of southern yellow-cheeked crested gibbons (Cambodia), northern buff cheeked/yellow-cheeked crested gibbons (Vietnam), and northern grey gibbons (Malaysia), along with randomly selected images from the 'Noise' category used in the present study.** The frequency range is 0.5 – 1.6 kHz for grey gibbons and 0.5 – 3.0 kHz for crested and buff cheeked gibbons. When images were input into the model, they did not have axis labels.

*Model architecture*

CNNs combine convolution and pooling into a sequence of layers; the number of layers is generally referred to as depth, and the overall depth and width of the model is generally thought to impact performance (Zeiler & Fergus, 2014). Here, we compare six commonly used CNN architectures, that vary in depth and composition. AlexNet was a pioneering CNN for image classification, and the introduction of AlexNet in 2012 led to substantial improvement in performance over existing methods. AlexNet contains five convolutional layers with three fully connected layers for a total of eight learned layers (Krizhevsky et al., 2017). Activation functions are used to transform the input signal to the output signal (Sharma et al., 2017), and AlexNet was one of the first to use Rectified Linear Unit (ReLu) activation function, which led to much faster training times (Sapijaszko & Mikhael, 2018; Figure 3). The VGG model architecture developed by the Visual Geometry Group (Simonyan & Zisserman, 2014) is much deeper than AlexNet; here we compare the VGG16 and VGG19 models which have 13 and 16 convolutional layers, respectively. An important difference between AlexNet and the VGG models is that the kernel size (or size of the convolutional filters) is smaller in the VGG models (Yu et al., 2016).

With traditional CNNs, there appears to be a maximum threshold of depth where the addition of more layers does not improve performance. He et al. (2016) noted that there is a 'degradation problem' that happens when networks get deeper, which leads to a decrease in performance that is not related to over-fitting. ResNets use residual connections (skip connections) to train very deep neural networks more efficiently, whereas the VGG relies on a simpler sequential architecture (Sapijaszko & Mikhael, 2018); this allows for ResNets to overcome the 'degradation problem'. Models in the ResNet family can be shallow or deep, and we chose to compare ResNet18, ResNet50, and ResNet152 which have 16, 48, and 150 convolutional layers, respectively, along with two fully connected layers. From a practical standpoint, CNNs with fewer layers are less computationally costly to train and deploy, which is why we wanted to compare both shallow and deep architectures, as this might mean they are more accessible for a higher number of practitioners who do not have access to large computing power.

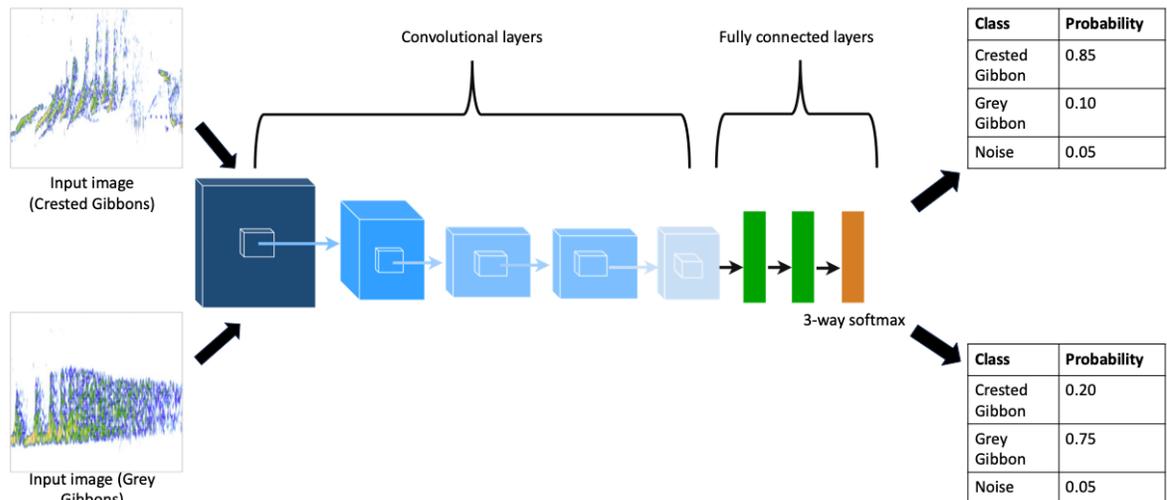

**Figure 3. A simplified overview of the AlexNET CNN architecture for our multiclass classification problem.** The CNN takes an image as an input, and then it passes through multiple convolutional layers (for AlexNet there are five convolutional layers; indicated in blue in the figure) which serve to provide a meaningful low dimensional representation of the image. The fully connected layers (green and orange) are used for classification, and in our multiclass problem with three classes (crested gibbons, grey gibbons, and noise) the final output layer (orange) has three dimensions which represent the three classes. This output can be converted to a confidence score that the model assigns to each class. For binary classification problems the output layer has one dimension that can be converted to a confidence of one of the two classes. The above image was adapted from (Yu et al., 2016), the

AlexNet architecture modified from (Krizhevsky et al., 2017), and the CNN was drawn using draw.io (https://www.drawio.com/).

*Model training and validation*

We took the training, validation, and test splits from different recorder locations, ensuring that we followed best practices for model evaluation (see Table 2 for sample size summary). We first converted the images to tensors, resized to 224x224 pixels, and applied mean and standard deviation normalization using the 'torchvision' R package (Falbel, 2022). We had many more noise samples than gibbon samples, however we left the dataset unbalanced, as this is more analogous to real-world scenarios of automated detection, and even with unbalanced datasets we achieved acceptable performance (see results). To account for this imbalance (Table 2), we used a weighted loss function that allowed us to assign weights to different classes (see below for details). All models were pretrained on the 'ImageNet' dataset (Deng et al., 2009) and are available in the 'torchvision' R package v 0.5.1 (Falbel, 2022). We allowed for early stopping and set 'patience=2' which specifies the number of epochs without improvement until the training is stopped.

We trained three sets of models for each of the model architectures, a binary model trained either on the 'grey gibbons' or 'crested gibbons' training datasets, and a multiclass model that was trained on both gibbon species along with the noise category. For all models, we set the maximum learning rate = 0.001 (Keydana, 2023), batch size = 32, and used an Adam optimizer (Kingma & Ba, 2014). We used the one-cycle learning rate strategy (Smith & Topin, 2018) implemented using the 'luz_callback_lr_scheduler' function in the 'luz' R package (Falbel, 2023). The one-cycle learning rate scheduler adjusts the learning rate dynamically by increasing it to a maximum value and then decreasing it throughout the training process (Paszke et al., 2019); one-cycle learning rate schedulers can be used in conjunction with adaptive gradient algorithms like the Adam optimizer to help accelerate convergence and improve generalization (Loshchilov & Hutter, 2019). For binary models, we used the 'nn_bce_with_logits_loss' loss function which combines a sigmoid layer with the binary cross entropy function, and specified the weight of the "Gibbon" class as 0.9 and the weight of the "Noise" class as 0.1. The use of a weighted loss function can be helpful in cases of class imbalance, as it makes the penalty for misclassification of the minority class higher (Lin et al., 2018). In this case, it means that misclassification of the gibbon samples is 9 times more costly than misclassification of noise samples. For the multiclass models we used 'nn_cross_entropy_loss' which computes the cross entropy loss between inputs and targets and is useful for multi-class classification problems (Falbel & Luraschi, 2023), and we specified the weight of the "Noise" class as 0.02, and the weights of the two "Gibbon" classes as 0.49.

**Table 2. Summary of training, validation, and test data sample size for initial model training and evaluation.**

| Data | Species | Training samples (Gibbon) | Training samples (Noise) | Validation samples (Gibbon) | Validation samples (Noise) | Test samples (Gibbon) | Test samples (Noise) |
|---|---|---|---|---|---|---|---|
| Danum Valley Conservation Training Data | Grey Gibbon | 349 | 1,651 | 78 | 206 | 76 | 332 |
| Jahoo Training Data | Crested Gibbon | 149 | 818 | 32 | 553 | 30 | 541 |

*Benchmarking Part 1: How variable are the results when running random iterations?*

The first part of our benchmarking approach aimed to investigate variation in performance metrics across model runs, and to compare performance metrics when fine-tuning the feature extractor compared to only retraining the classification head. Previous work found that when using > 25 samples, fine-tuning the feature extractor layer along with the output layer resulted in better performance (Dufourq et al., 2022). To evaluate the effect of fine-tuning on model performance in our system, we used two training approaches: (1) by setting requires_grad = TRUE, allowing the entire network to be trained, and (2) setting requires_grad = FALSE, where only the classification head was trained. Due to computational constraints of multiple model runs, we set models to run for 1 epoch over 3 iterations for each model architecture in our study: AlexNet, VGG16, VGG19, ResNet18, ResNet50, and ResNet152 (see training details below).

*Benchmarking Part 2: Which combination of architecture and epochs lead to best performance?*

The second part of our benchmarking approach compared the impact of varying the number of epochs on model performance. For this section, we trained AlexNet, VGG16, VGG19, ResNet18, ResNet50, and ResNet152 models that were allowed to run for 1, 2, 3, 4, 5, and 20 epochs. We allowed for early stopping, with 'patience=2' which indicates the number of epochs without improvement until the training is stopped. For this analysis, we included only fine-tuned models, wherein the feature extractor was fine-tuned in addition to the classification layer.

*Benchmarking Part 3: Does data augmentation improve model performance?*

We initially aimed to use the test data from the original split of training data (Table 2) to identify the best-performing combinations of CNN architecture, number of epochs, and type of training data (binary or multiclass) to use for further model evaluation. However, we found that multiple model configurations led to similar performance, with high maximum F1 score on the test data (see results below) which is indicative of overfitting. To address this issue, we conducted data augmentation experiments. For the "noise added" training datasets we used the 'noise' function in the 'tuneR' package (Ligges et al., 2016) to add white noise and pink noise to each training clip. For this set, we created two new sound files for each training clip. For the "cropped" dataset, we created two randomly cropped files for each training clip. To do this, we randomly selected both a start time within the training clip and a random clip duration that was less than the actual duration of the clip, and we cropped the audio and then normalized using the 'normalize' function in 'tuneR'. For the "duplicated" dataset, we copied the training images five times, so that each sample was represented multiple times in the training data. When importing the image data for model training, the 'torchvision' package (Falbel, 2022) provides the 'transform_color_jitter' function to randomly change the brightness, contrast, and saturation of an image. For both the original training data and the "copy" data, we evaluated model performance with and without color jitter. For this set of experiments, we focused on two model architectures, AlexNet and ResNet50. We chose AlexNet to serve as a baseline, as it is the least complex architecture, and ResNet50 as it exhibited consistently high performance in the earlier benchmarking experiments but is not as computationally costly as ResNet152. We ran these experiments with models trained using 1 and 5 epochs.

*Benchmarking Part 4: Comparison with BirdNET*

We used BirdNET V2.4, which uses an EfficientNetB0-like backbone (Tan & Le, 2020) and is trained predominantly on labeled acoustic data from Xeno-canto (https://xeno-canto.org/) and the Macaulay Library (https://www.macaulaylibrary.org/). We used the command line interface to implement the 'train.py' function in BirdNET (https://github.com/kahst/BirdNET-Analyzer) for each training dataset (binary grey gibbons, binary crested gibbons, and a multiclass combined model). This function allows users to train a new classifier for sound types not currently in BirdNET. To create predictions from the trained models, we used the 'analyze.py' function. This version of BirdNET returns predictions for 3-sec windows. For both training and analysis, we used the default settings apart from setting fmin = 500 Hz and fmax = 2,000 Hz for grey gibbons, and fmin = 500 Hz and fmax = 3,000 Hz for crested gibbons. For the multiclass models, we trained two separate models setting fmax=2000 for grey gibbons or fmax=3000 for crested gibbons. We used default settings, where the number of training epochs was 50, the batch size was 32, the validation split ratio was 0.2 and the learning rate was 0.001. We utilized a single-layer classifier with no hidden units and did not apply dropout or mixup. We deployed the trained models over the test datasets for generalizability from different sites. To evaluate performance, within each test clip, we identified the 3-s clip with the highest confidence score.

*Performance and deployment*

To evaluate performance for the binary 'torch for R' models, the output layer had a single dimension, so we used a sigmoid function to convert the output of the CNN predictions on the test datasets to values between 0 and 1. For the multiclass models, the output layer had three dimensions, representing the three classes in our data, and we computed class probabilities by applying a softmax function. For all benchmarking and performance evaluations, we report maximum F1 score estimated over a range of thresholds at 0.1 increments between 0.1-1.0 calculated using the 'caret' R package (Kuhn, 2008), and we report the Area Under the Receiver Operating Characteristic Curve (AUC-ROC), which is a threshold-independent score and does not require threshold tuning. We used the 'ROCR' R package to calculate AUC-ROC (Sing et al., 2005). For the multiclass models we used the 'one-vs-all' approach to calculate AUC-ROC, where AUC-ROC is calculated for each class, and report the AUC-ROC for the class of interest. We also report the false positive rates (FPR) for the wide arrays at Danum Valley and Jahoo, as we deployed the model over the longer sound files which would allow us to calculate this more effectively. We used the 'confusionMatrix' function in the 'caret' package (Kuhn, 2008) to calculate specificity, and FPR was derived using the standard formula, FPR = 1-specificity.

*Model deployment*

We deployed trained models over the wide arrays at Danum Valley and Jahoo. For Danum Valley we focused on 06:00-08:00 LT (Clink et al., 2020) and for Jahoo we focused on 05:00- 06:00 LT (Clink et al., 2024), as these are the peak calling times at these two sites. The model outputs spectrogram images of detections, and a single observer (DJC) manually sorted the images into either true positives (grey gibbon or crested gibbon) or false positives (noise). For cases of uncertainty, we also listened to the sound clip to assist with manual sorting into correct categories. We then created a call event density map using inverse distance weighed interpolation (Wrege et al., 2017) with the R package 'gstat' (Pebesma, 2004). We standardized the number of calling events by the number of hours analyzed for each recorder location. Recorders were deployed on a grid at approximately 750-m spacing for Danum Valley Conservation Area, Malaysia and approximately 2-km spacing for Jahoo, Cambodia. Using an Apple M2 Max with 12 core CPU, 30 core GPU, and 32 GB memory we processed 1-hour files at 32 kHz sample rate or 2-hour files at 16 kHz sample rate in approximately 3 minutes. This estimate includes the image creation step, which can add to the processing time. All analyses were done using R programming language v 4.2.1 (R Core Team, 2022) using the 'gibbonNetR' R package (v1.0.0) (Clink & Ahmad, 2024).

*Data availability*

Scripts needed to run the analyses are located at: https://github.com/DenaJGibbon/torch-for-R-gibbons. Data needed to reproduce analyses are available on Zenodo: https://doi.org/10.5281/zenodo.10948975.

**Results:**

*Benchmarking Part 1*

We trained each model architecture for 1 epoch over three replicate runs, and found that the maximum F1 score (calculated across evenly spaced 0.1 threshold increments between 0.1 - 1.0) was generally higher for models with fine-tuning (Figure 4). Both maximum F1 and the standard error varied by model type and class. We also found that AUC-ROC was generally higher for models that were fine-tuned, however there was variation across model

architectures. Standard error in AUC-ROC across runs was generally small, except for the VGG16

and VGG19 architectures (Figure 4).

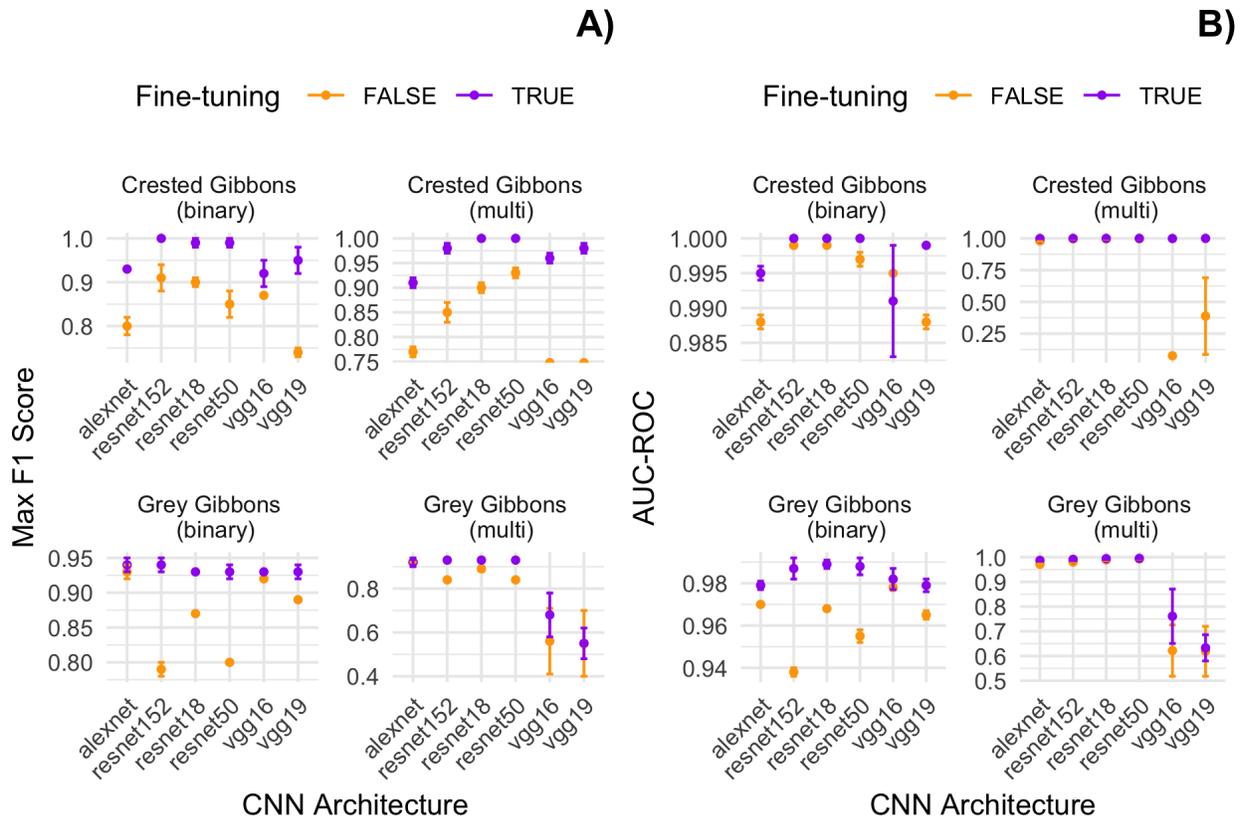

Figure 4. A comparison of the maximum F1 score ± standard error (A) and maximum AUC-ROC ± standard error (B) over three replicate model runs for each of the model architectures in our analysis. The x-axis indicates the CNN architecture, and the color indicates whether the model feature extractor was fine-tuned (purple) or not (orange). The y-axis indicates the value for each metric. Note the variation in y-axis range for each data type.

*Benchmarking Part 2*

For crested gibbons, many combinations of CNN architectures and the number of epochs resulted in the same performance when training binary classification models (F1 score =

1; AUC-ROC=1). For grey gibbons, the AlexNet architecture trained for 4 epochs (with early stopping) had the highest performance for the binary classification model (F1 score = 0.95; AUC-ROC = 0.99). For the multi-class classification models, we found that for crested gibbons many model combinations had high maximum F1 score = 1, and AUC-ROC =1. For and the RestNet50 model trained for 20 epochs with early stopping had the highest maximum F1 score (0.96) (see Appendix Table 1).

*Benchmarking Part 3*

We evaluated model performance using multiple types of data augmentation, including adding noise, random cropping, and adding color jitter to the spectrogram image on the original dataset or a duplicated dataset. We report both AUC-ROC and maximum F1 score. We found that on the original test data split, performance was similar across data augmentation and model types, however the multiclass models had consistently higher performance (Appendix Figure 1). To further evaluate the impact of data augmentation, we deployed the multiclass models trained on the different augmented training datasets over test datasets from Maliau Basin Conservation Area, Malaysia (grey gibbons) and Dakrong Nature Reserve, Vietnam (crested gibbons). We found that the "duplicated" dataset with color jitter had consistently high performance on the test dataset (Figure 5), and both architectures exhibited relatively high performance.

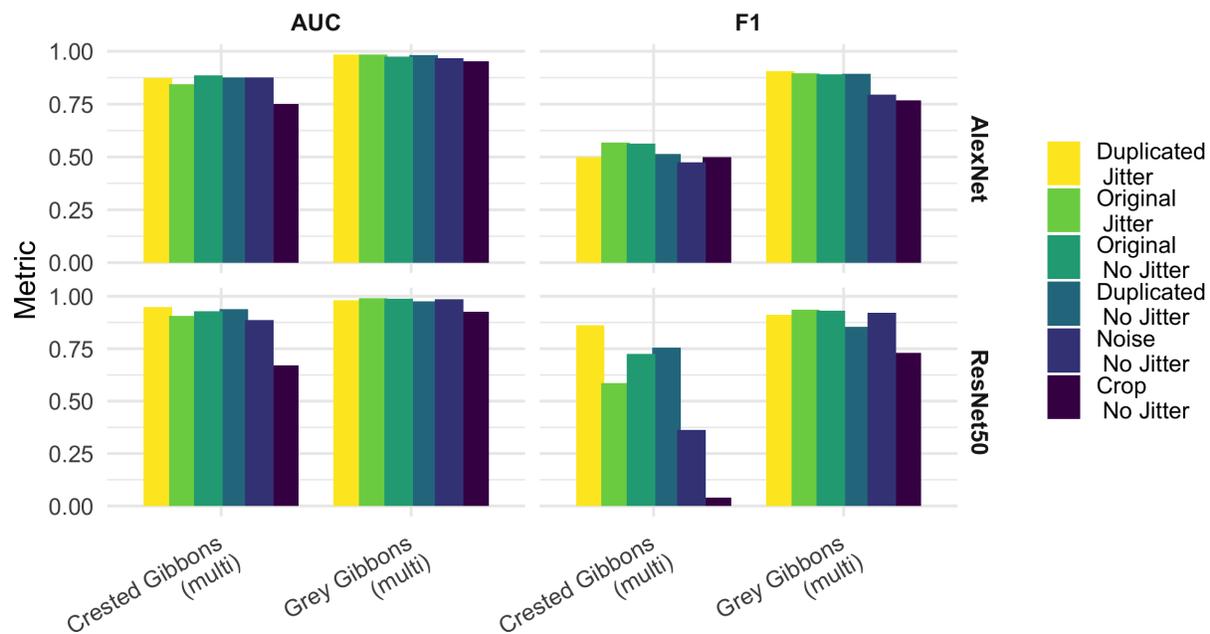

Figure 5. Results of data augmentation benchmarking experiments for multiclass models on test datasets from Maliau Basin Conservation Area, Malaysia (grey gibbons) and Dakrong Nature Reserve, Vietnam (crested gibbons). The metrics show the highest AUC-ROC and maximum F1-score for each type of data augmentation approach for both AlexNet and ResNet50 architectures.

*Benchmarking Part 4: Comparison with BirdNET*

For benchmarking with BirdNET, we used the AlexNet and ResNet50 models trained on the "duplicated" dataset with color jitter deployed over the test datasets from Maliau Basin Conservation Area, Malaysia (grey gibbons) and Dakrong Nature Reserve, Vietnam (crested gibbons) for comparison. We deployed binary and multiclass BirdNET models over the same test datasets. We found comparable performance across all models, apart from the BirdNET multiclass had lower performance on the crested gibbon test dataset (Table 4; Appendix Figure 2). For a summary of benchmarking objectives, results, and key takeaways see Table 5.

**Table 4. A comparison AlexNet and ResNet50 architecture performance to that of BirdNET binary and multiclass models.** The table shows the maximum F1 score, precision, recall, threshold, and AUC for each model configuration. The best performing model for each species as determined by AUC-ROC is indicated in bold.

| Species | Model | Max_F1 | Precision | Recall | Threshold | AUC-ROC |
|---|---|---|---|---|---|---|
| Crested Gibbon | **ResNet50 Multi** | **0.86** | **1.00** | **0.76** | **0.1** | **0.95** |
| | BirdNET Binary | 0.87 | 0.93 | 0.81 | 0.1 | 0.90 |
| | BirdNET Multi | 0.34 | 0.36 | 0.32 | 0.1 | 0.61 |
| Grey Gibbon | **AlexNet Multi** | **0.90** | **0.87** | **0.94** | **0.7** | **0.98** |
| | BirdNET Multi | 0.81 | 0.73 | 0.91 | 0.9 | 0.93 |
| | BirdNET Binary | 0.86 | 0.81 | 0.91 | 0.4 | 0.91 |

Table 5. A summary of benchmarking experiments, objectives, and key findings for the present study.

| Benchmarking | Objective | Methods Summary | Models/ Configurations | Results Summary | Key Takeaways |
|---|---|---|---|---|---|
| *Part 1: How variable are the results when running random iterations?* | Investigate variation in performance metrics across replicate model runs and compare fine-tuning approaches. | Two training approaches: (1) Fine-tuning entire network; (2) Only retraining classification head.<br><br>Models run for 1 epoch over 3 iterations. | AlexNet, VGG16, VGG19, ResNet18, ResNet50, ResNet152 | Fine-tuning generally led to better performance.<br><br>ResNet archtectures had more consistent performance. | Focus only on fine-tuned models. |
| *Part 2: Which combination of architecture and epochs lead to best performance?* | Identify best architecture and number of epochs for model performance. | Training models for 1, 2, 3, 4, 5, and 20 epochs with early stopping (patience=2).<br><br>Only fine-tuned models. | AlexNet, VGG16, VGG19, ResNet18, ResNet50, ResNet152 | Many different model combinations led to high performance.<br><br>Very high performance metrics indicative of possible overfitting to training data.<br><br>AlexNet and ResNet50 consistently high performers. | Focus on AlexNet and ResNet50. |

| Part 3: Does data augmentation improve model performance? | Assess impact of data augmentation on performance. | Added noise (white/pink), cropped, or duplicated training data.  Evaluated with and without color jitter.  Models trained for 1 and 5 epochs. | AlexNet (baseline), ResNet50 (high performance) | Multiclass ResNet50 models with copied training data and color jitter are best performers. | Copied training data with color jitter improves model performance.  Multiclass models perform slightly better. |
|---|---|---|---|---|---|
| Part 4: Comparison with BirdNET | Compare performance with BirdNET V2.4. | Trained BirdNET using binary and multiclass datasets.  Used the test dataset from different sites to test generalizability. | AlexNet, ResNet50, BirdNET V2.4 | 'Torch for R' models slightly outperformed BirdNET models, especially for multiclass crested gibbons. | Performance across model types varies but is comparable. |

*Final performance evaluation*

To report the final performance of the models, we used the ResNet50 multi-class model trained on the "duplicated" dataset with color jitter for both species, and ran them over test files from the wide arrays at each site (see methods for details). We found that both models achieved acceptable performance (Table 6; Figure 6 and that false positive rates were generally low.

**Table 6. Model performance for test files from wide arrays at Danum Valley Conservation Area, Malaysia, and Jahoo, Cambodia.** Models were multiclass ResNet50 architecture trained on the "duplicated" dataset with color jitter. AUC-ROC is shown once for each class as it is threshold independent.

| Type | Class | F1 | AUC-ROC | Threshold | Precision | Recall | FPR |
|---|---|---|---|---|---|---|---|
| Best F1 | Crested Gibbons | 0.82 | 0.95 | 0.4 | 0.89 | 0.75 | 0.005 |
| Best F1 | Grey Gibbons | 0.78 | 0.93 | 0.2 | 0.83 | 0.73 | 0.020 |
| Best Precision | Crested Gibbons | 0.75 | ~ | 0.9 | 0.97 | 0.60 | 0.001 |
| Best Precision | Grey Gibbons | 0.66 | ~ | 0.9 | 0.95 | 0.51 | 0.004 |
| Best Recall | Crested Gibbons | 0.79 | ~ | 0.1 | 0.75 | 0.84 | 0.014 |
| Best Recall | Grey Gibbons | 0.76 | ~ | 0.1 | 0.75 | 0.77 | 0.034 |

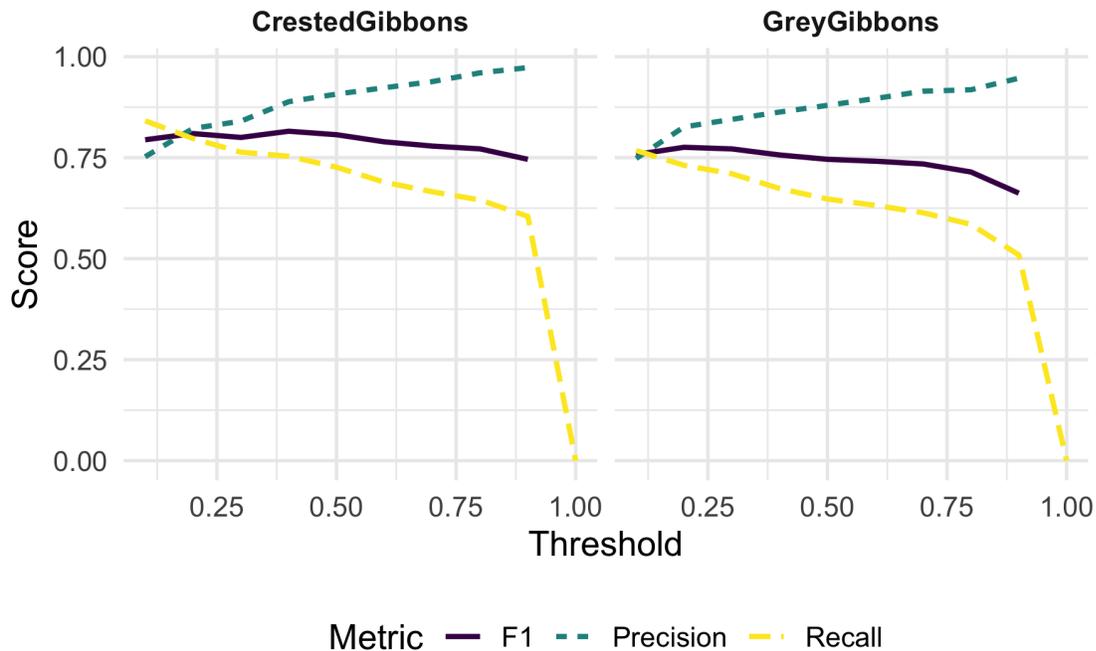

**Figure 6. Precision, recall, and F1 score as a function of confidence score for test datasets from the wide array for crested and grey gibbons.** Models used were multiclass ResNet50 architecture trained on the "duplicated" dataset with color jitter.

*Deploying the models*

We deployed the ResNet50 multi-class model trained on the "duplicated" dataset with color jitter over wide PAM arrays in Danum Valley Conservation Area, Malaysia, and Jahoo, Cambodia, setting the confidence threshold = 0.9. For each site, we manually assigned all detections as either true positive (gibbon) or false positive (noise). We found that 285 out of 3,046 were false positives for Danum Valley, resulting in a precision of ~ 0.90. For Jahoo, we found that 203 out of 1,919 were false positives, resulting in a precision of ~0.89. It is important to note that we could not effectively estimate recall for the large-scale array, and it is likely that recall is different from that we report on our test data. We found that there were substantial differences in the number of calls detected across ARU locations (Figure 7). We found that in some cases the pattern matching ability of the algorithms outperformed visual pattern matching of human observers, as we had to listen to the corresponding clip to determine if the signal was a true or false positive. See Figure 8 for examples of true positive spectrogram images across different confidence scores.

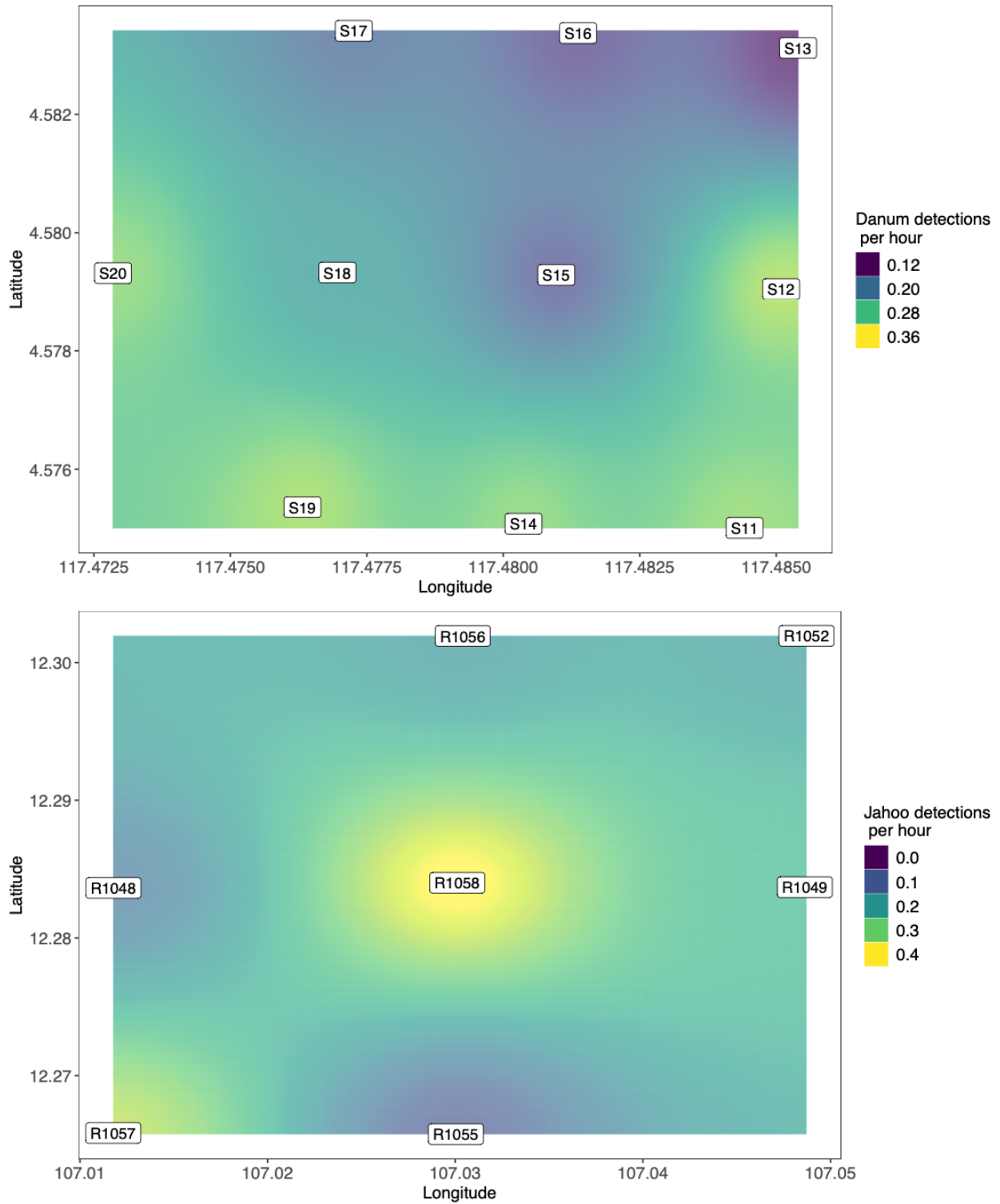

**Figure 7. Call event density for Danum Valley Conservation Area, Malaysia (top) and Jahoo, Cambodia (bottom).** Recorders were placed at ~750-m spacing in Danum Valley and ~2-km at Jahoo. The number of detections was standardized by the number of recording hours at each recorder location.

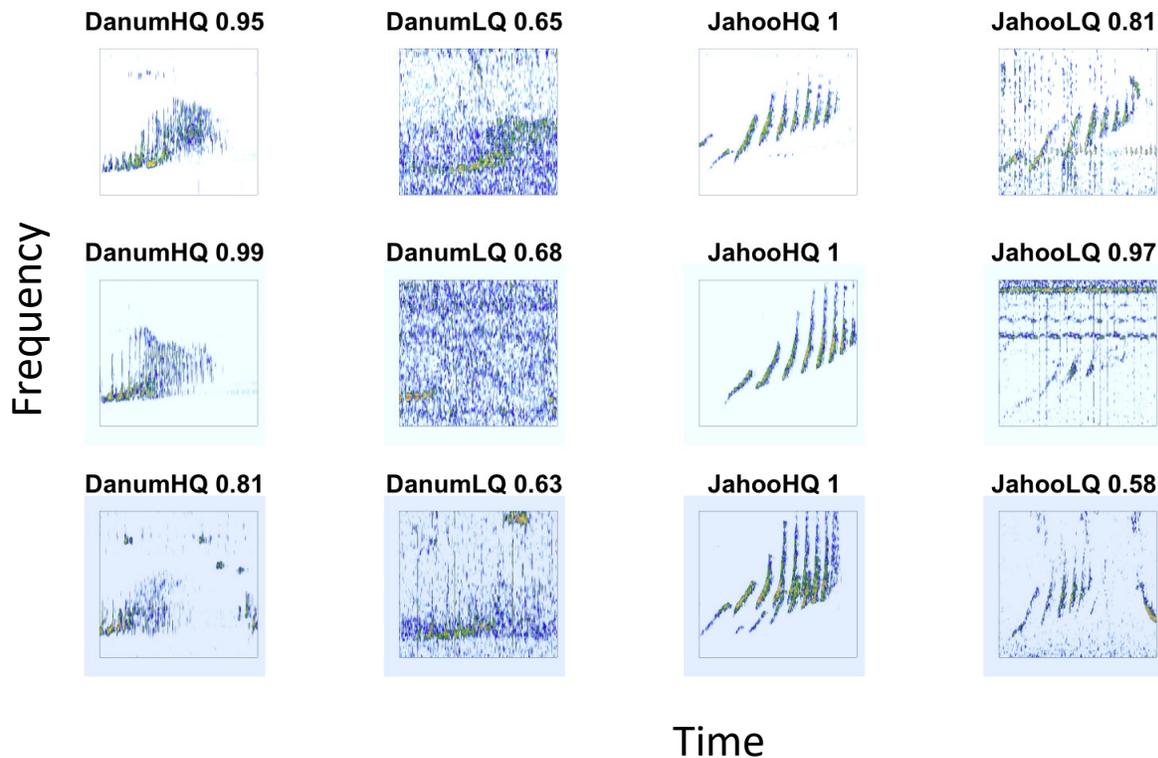

**Figure 8. Spectrogram images of detections from the Danum Valley Conservation Area, Malaysia wide array (grey gibbons), and from the Jahoo, Cambodia wide array (crested gibbons).** Images are divided into high-quality (HQ) and low-quality (LQ) subjective categories and names include the corresponding confidence as assigned by the CNN. The low-quality images required subsequent aural verification by a human observer to ensure they were classified correctly.

**Discussion:**

Here, we provide a benchmark for automated detection of gibbon sounds using CNNs in the 'torch for R' environment. To our knowledge this is the first implementation of automated detection from PAM data using the "torch for R" framework (Falbel & Luraschi, 2023; Keydana, 2023). We compared the performance of different architectures when trained for a binary classification problem (gibbons vs noise) or trained for a multi-class classification problem (training data included two species of gibbons). Although the two gibbon species are not sympatric, we predicted that including more samples of gibbon calls regardless of the species would improve performance. We did not provide a comprehensive analysis of all available architectures and hyperparameters but found acceptable performance with the subset that we examined.  Similar to other studies, we found that the best performing architecture was dependent on the signal type (e.g. species of gibbon), however, ResNet architectures had consistently high performance. We found that the performance of the multi-class models was generally higher, which we attribute to reduced overfitting to the training data given the higher number of samples included. Our benchmarking results revealed that color jitter was an effective data augmentation strategy for our dataset, and that fine-tuning the entire model, instead of just retraining the classification layers led to better performance. We were able to process 1-hour sound files with a 32 kHz sample rate in approximately 3 minutes, which substantially reduces processing time relative to human scanning of spectrograms.

Our initial evaluation of model performance used training, validation, and test datasets from different ARU locations within a particular site. Although we worked to ensure that there was no data leakage between the training and test clips, the performance values we report from these initial tests are much higher than subsequent evaluations on datasets from different PAM arrays. This highlights the importance of using different test datasets to report the generalizability of trained models (Stowell, 2022). We believe the initial high performance may be indicative of overfitting due to a lack of diversity in acoustic signals contained in the training and test datasets, along with a smaller number of noise samples relative to real-world deployments for automated detection where a substantial amount of the clips are noise. However, the top performing models also had acceptable performance on test datasets from other sites and gibbon species indicating that this is an effective approach to identify combinations of CNN architectures and hyperparameters for subsequent testing.

Humans are good at pattern recognition, and tend to learn faster (e.g. with fewer training samples) than machine learning algorithms (Kühl et al., 2022). In our study, there were cases where the algorithm outperformed the human observer in assigning detections to true positives, as we had to rely on aural confirmation to verify detections were true positives. Overall, we found that the approaches presented here led to acceptable performance, and slight improvements from traditional machine learning automated detection methods for grey gibbons from Danum Valley Conservation Area (F1 score < 0.8 ) (Clink et al., 2023). We found that our results were comparable, albeit slightly lower, to those found for gibbons using a similar transfer learning approach (Dufourq et al., 2022). We believe the slightly lower performance in our system may be due to the fact we used test data sets from different PAM arrays than the training data. Interestingly, we also found that our method was comparable to and in some cases slightly outperformed BirdNET, which is considered a state-of-the-art model for bioacoustics applications (Ghani et al., 2023). One possible reason is that the current version of BirdNET (v2.4) works only on 3-s clips, and due to this limitation, it may not capture longer temporal patterns of variation in gibbon calls that the 'torch for R' models were able to capture due to the longer window size (12-s). An important clarification is that the BirdNET models did not use any form of data augmentation, whereas the 'torch for R' models did, so it is possible that data augmentation would lead to superior performance of BirdNET.

We found that the range of number of detections per hour in Danum Valley and Jahoo were comparable, although there were clear differences across recording locations within a single site. Gibbon female calls often occur within a duet bout wherein there are multiple replicates of the great call. The number of great calls and the duration of the duets can vary across pairs and it has been shown in cao vit gibbons (*Nomascus nasutus*) the number of great calls emitted in a duet is influenced by pair bond strength (Ma et al., 2022). Therefore, it is possible that the differences we see within sites are due to differences of individual gibbon females recorded at particular ARUs. Duetting in gibbons is density-dependent, and can also be influenced by outside factors such as rainfall (Clink et al., 2020; Cowlishaw, 1992). Territory size of both species is not well-known, and it is unclear how much territory overlap there is between neighboring groups. However, even if there are differences in home range size across species, this did not impact the number of calls detected per hour.

The field of automated detection is rapidly evolving, and we provide these results as a benchmark for future automated detection approaches. Gibbons tend to be highly vocal, which means that collating training datasets is easier than for less vocal animals. However, future work investigating the impact of other types of data augmentation (e.g. sound synthesis (Guei et al., 2024)) or modification of the number of training samples on model performance will be informative (Dufourq et al., 2022; Nolasco et al., 2023). In addition, it is possible that a global model trained on all 20 gibbon species may outperform other types of models. Therefore, we propose that future work combines multiple gibbon species into a training dataset and trains a deep learning model from scratch. However, in contrast to avian communities in tropical forests, there are generally only one to two gibbon species at a particular site, so it is possible that a simple binary classifier maybe sufficient.

The R programming environment is one of the commonly used languages for ecologists (Scavetta & Angelov, 2021), making the 'torch for R' environment a potentially accessible deep learning tool for this community. However, a proficiency in coding is still needed to effectively implement the tools in their current form. We found the documentation for 'torch for R' to be comprehensive, and it includes step-by-step tutorials for both image classification and audio classification (Keydana, 2023). We initially aim to adapt the audio classification approach, but found that the training time for models on our system was prohibitively long. In future work we would like to adapt the approach we present here so that it uses spectrograms as inputs to the model instead of images. The current version of the 'gibbonNetR' package v1.0.0 (Clink & Ahmad, 2024) that we used for this analysis does not allow for processing on multiple threads, which means that it takes longer than BirdNET to process multiple sound files or large PAM deployments. BirdNET v2.4, while faster to train and deploy, has a few limitations including a fixed window size of 3-s and subpar performance on acoustic signals that are at infra- or ultra-sonic signals. Therefore, the optimal tool will therefore depend on the specific requirements of the user and the characteristics of the target species' vocal behavior.

As the majority of gibbon species are endangered, and all of them emit loud calls that can be heard several kilometers away (Geissmann, 2002), they are an excellent model for developing PAM and automated detection approaches. In addition to developing effective models for automated detection, more work needs to be done to make these approaches accessible to conservation practitioners. One promising approach for making automated detection using deep learning accessible to non-coders is that found in the BirdNET GUI (https://github.com/kahst/BirdNET-Analyzer). In addition, implementing transfer learning in the

'torch for R' environment into a shiny app (similar to (Ruff et al., 2020)) will make the approach presented here more accessible.

## Acknowledgements

DJC acknowledges the Fulbright ASEAN Research Award for U.S. Scholars for providing funding for the field research in Danum Valley Conservation Area, Malaysia. We thank the Royal Government of Cambodia and the KSWS REDD+ program for providing funding for field work in Jahoo, Cambodia. We thank Kyle Burrichter for his assistance with field data collection at Jahoo. TVT thanks Dakrong Nature Reserve for permission to conduct the field survey in Vietnam.

## Author contributions

*DJC* and *HK* conceived the ideas and designed methodology; *MH, RS, HB, CA, TVT, HNT, and TNT* collected the data; *JK, HCJ* and *DJC* analysed the data; *DJC, JK, and HCJ* led the writing of the manuscript. *AHA* provided logistical and in-country and contributed to the ideas and methodology. All authors contributed critically to the drafts and gave final approval for publication.

## Conflict of interest

The authors have declared that no competing interests exist.

## Ethics statements

**Appendix**

**Appendix Table 1. Results of the "Benchmarking Part 2" experiments comparing the performance of different CNN architectures trained for 1, 2, 3, 4, 5 and 20 epochs with early stopping.** For crested gibbons there were multiple combinations with high performance.

| Species | N epochs | CNN Architecture | Threshold | Precision | Recall | F1 | AUC |
|---|---|---|---|---|---|---|---|
| Crested Gibbon binary | 1, 3, 4, 5 | resnet152 | 0.1, 0.2, 0.3, 0.4, 0.5, 0.6, 0.7, 0.8, 0.9 | 1.00 | 1.00 | 1.00 | 1.00 |
|  | 2, 20 | vgg19 | 0.4, 0.5, 0.6, 0.7, 0.8 | 1.00 | 1.00 | 1.00 | 1.00 |
|  | 3, 4, 5 | resnet18 | 0.1, 0.2, 0.3, 0.4, 0.5, 0.6, 0.7 | 1.00 | 1.00 | 1.00 | 1.00 |
|  | 3 | resnet50 | 0.4 | 1.00 | 1.00 | 1.00 | 1.00 |
|  | 5 | vgg16 | 0.9 | 1.00 | 1.00 | 1.00 | 1.00 |
| Grey Gibbon binary | 4 | alexnet | 0.9 | 0.97 | 0.93 | 0.95 | 0.99 |
|  | 1, 2, 3, 4, 5, 20 | resnet18 | 0.1, 0.2, 0.3, 0.4, 0.5, 0.6, | 1.00 | 1.00 | 1.00 | 1.00 |

| Species | N epochs | CNN Architecture | Threshold | Precision | Recall | F1 | AUC |
|---|---|---|---|---|---|---|---|
| Crested Gibbon multi | 1, 3, 4, 5 | resnet50 | 0.7, 0.8, 0.9 0.1, 0.2, 0.3, 0.4, 0.5, 0.6, 0.7, 0.8, 0.9 | 1.00 | 1.00 | 1.00 | 1.00 |
| | 4, 20 | resnet152 | 0.1, 0.2, 0.3, 0.4, 0.5, 0.6, 0.7, 0.8, 0.9 | 1.00 | 1.00 | 1.00 | 1.00 |
| | 4, 5, 20 | vgg16 | 0.1, 0.2, 0.3, 0.4, 0.5, 0.6, 0.7, 0.8, 0.9 | 1.00 | 1.00 | 1.00 | 1.00 |
| | 4, 5, 20 | vgg19 | 0.2, 0.3, 0.4, 0.5, 0.6, 0.7, 0.8, 0.9 | 1.00 | 1.00 | 1.00 | 1.00 |
| | 4 | alexnet | 0.4 | 1.00 | 1.00 | 1.00 | 1.00 |
| Grey Gibbon multi | 2 | resnet50 | 0.6 | 0.95 | 0.96 | 0.95 | 0.99 |

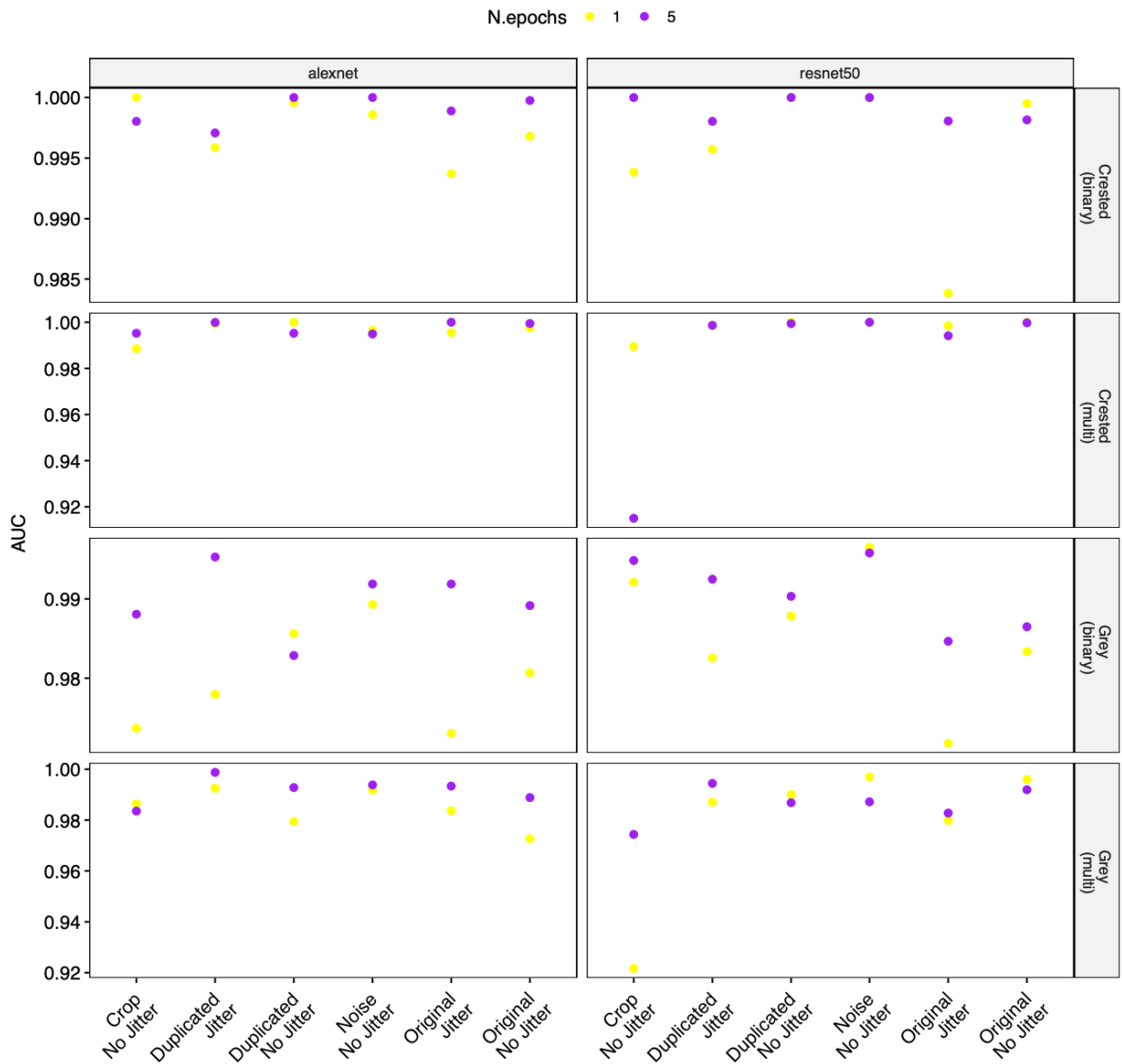

**Appendix Figure 1. AUC–ROC scores across training data augmentation types and architectures for "Benchmarking Part 3".** AUC–ROC was calculated one the original test data split. The color of the points indicates if the model was trained for one epoch (yellow) or five epochs (purple).

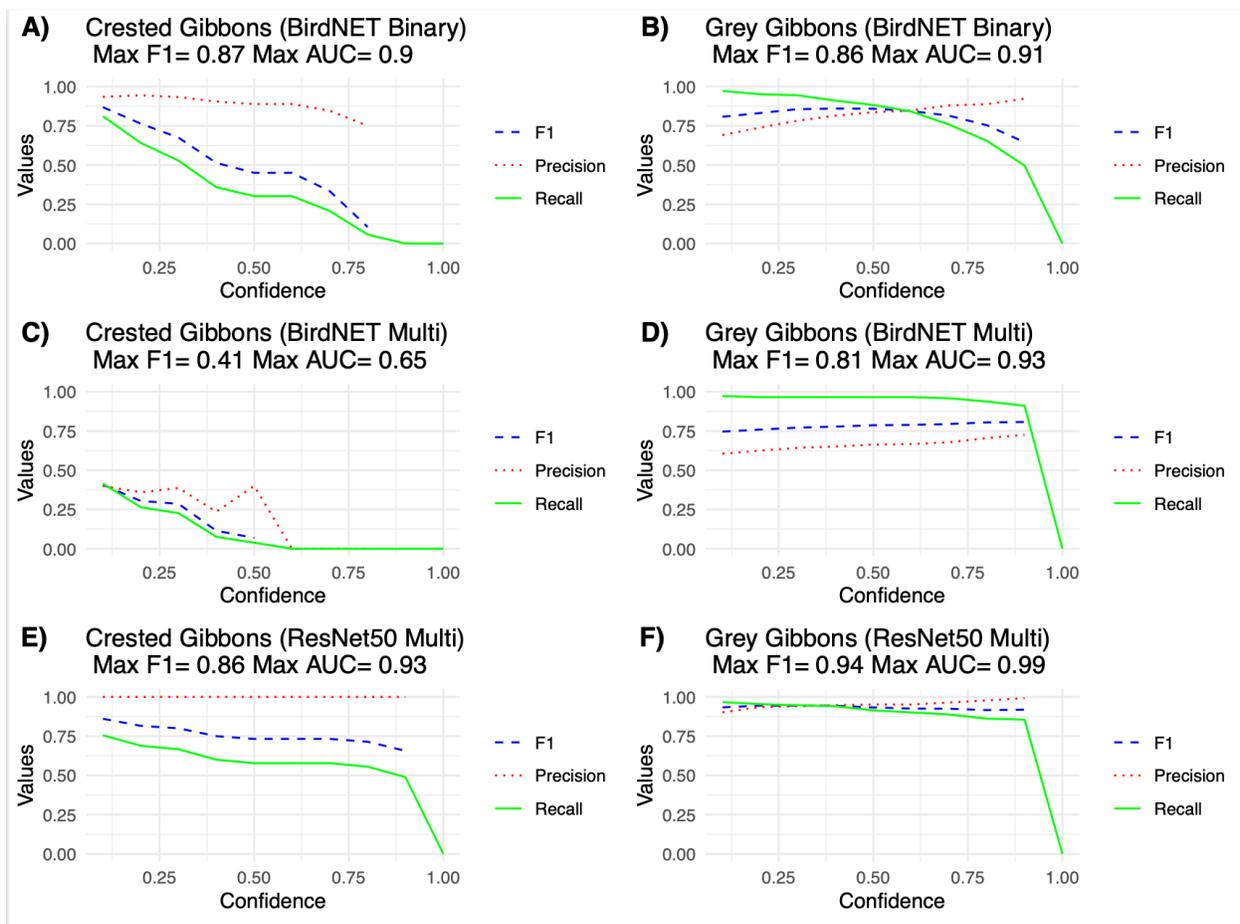

Appendix Figure 2. Precision, recall, and F1 score as a function of confidence score for BirdNET (binary and multiclass) and ResNet50 (multiclass only) models for classification of gibbon calls. BirdNET models were trained on the original training data, and multiclass ResNet50 models were trained on the "duplicated" dataset with color jitter.